\documentclass[aps,preprintnumbers,superscriptaddress,showpacs]{revtex4}
\usepackage{epsfig}
\usepackage{psfrag}
\usepackage{amsfonts}
\usepackage{graphicx}
\usepackage{dcolumn}
\usepackage{bm}

\begin{document}

\title{Jet quenching parameter from a soft wall AdS/QCD model}

\author{Xiangrong Zhu}
\email{xrongzhu@zjhu.edu.cn} \affiliation{School of Science,
Huzhou University, Huzhou 313000, China}

\author{Zi-qiang Zhang}
\email{zhangzq@cug.edu.cn} \affiliation{School of Mathematics and
Physics, China University of Geosciences, Wuhan 430074, China}

%%%%%%%%%%%%%%%%%%%%%%%%%%%%%%%%%%%%%%%%
\begin{abstract}
We study the effect of chemical potential and nonconformality on
the jet quenching parameter in a holographic QCD model with
conformal invariance broken by a background dilaton. It turns out
that the presence of chemical potential and nonconformality both
increase the jet quenching parameter thus enhancing the energy
loss, consistently with the findings of the drag force.
\end{abstract}

\maketitle
%%%%%%%%%%%%%%%%%%%%%%%%%%%%%%%%%%%%%%%%
\section{Introduction}

It is believed that the high energy heavy-ion collisions at both
the Relativistic Heavy Ion Collider (RHIC) and the Large Hadron
Collider (LHC) have produced a new type of matter so-called quark
gluon plasma (QGP) \cite{EV,KA,JA}. One of the key characteristics
of QGP is jet quenching: when high energy partons go through the
thermal medium, they will interact with the medium and then lose
their energy via collisional and radiative processes. This
phenomenon is usually characterized by the jet quenching parameter
$\hat{q}$, defined as the averaged transverse momentum broadening
squared per unit mean free path \cite{XN1,RB}. For the study of
$\hat{q}$ in perturbative QCD, see \cite{XN2,XN3}. However, lots
of experiments indicate that QGP is a strongly coupled medium.
Thus, it is of interest to study the jet quenching in strongly
coupled settings.

AdS/CFT \cite{Maldacena:1997re,MadalcenaReview,Gubser:1998bc}, a
conjectured duality between a type IIB string theory in
$AdS_5\times S^5$ and $\mathcal{N}=4$ Super Yang-Mills (SYM) in
(3+1)-dimensions, provides a powerful tool to describe strongly
coupled gauge theories. Over the past two decades, this duality
has yielded many important insights for studying various aspects
of QGP (see \cite{JCA,AAD} for recent reviews with many
phenomenological applications). Using AdS/CFT, H. Liu et. al,
proposed a nonpeturbative definition of $\hat q$, based on the
computation of light-like adjoint Wilson loops, and then applied
to calculate the jet quenching parameter for $\mathcal{N}=4$ SYM
plasma at finite temperature \cite{liu,liu0}. Since then, this
idea has been extended to various holographic models. For
instance, the finite 't Hooft coupling corrections on $\hat{q}$
are studied in \cite{NA,ZQ,JSA0,KB}. The effect of chemical
potential on $\hat{q}$ is discussed in \cite{FL,SD}. The effect of
electric or magnetic field on $\hat{q}$ appeared in
\cite{JSA,JSA1,ZQ1}. Also, this quantity has been investigated in
some nonconformal settings \cite{DL,UG,RR}. Other interesting
results can be found in \cite{MC,DG,SL,AF1,AB1,MB,EC,SHI,EN,AS}.

In this paper, we reexamine the jet quenching parameter in a soft
wall AdS/QCD model, motivated by the soft wall model of
\cite{AKE}. Especially, we adopt the SW$_{T,\mu}$ model by P.
Colangelo et. al, \cite{PCO} which was applied to investigate the
free energy of a heavy quark antiquark pair and the QCD phase
diagram. It is found that such a model provides a well
phenomenological description of quark-antiquark interaction. Also,
the resulting deconfinement line in the $\mu - T$ (with $\mu$ the
chemical potential and $T$ the temperature) plane is similar to
that obtained by lattice and effective models of QCD.
Subsequently, the authors of Ref. \cite{zq} studied the imaginary
part of heavy quark potential in the SW$_{T,\mu}$ model and found
the inclusion of nonconformality reduces the quarkonia
dissociation, reverse to the effect of chemical potential. More
recently, the drag force \cite{YH} has been discussed in the same
model and the results show that the presence of nonconformality
and chemical potential both enhance the drag force. Further
studies of models of this type, see \cite{OA1,CPA,PCO1,CE,XCH}.
Inspired by this, we want to study the jet quenching parameter in
the SW$_{T,\mu}$ model. Specifically, we want to understand how
nonconformality and chemical potential modify this parameter,
respectively. Also, we will compare our results with that of
\cite{YH} and to see whether nonconformality and chemical
potential have the same effect on the energy loss of heavy quarks
(related to the drag force) as with light quarks (associated with
the jet quenching parameter)? It is the purpose of the present
work.

The organization of the paper is as follows. In section 2, we
briefly review the SW$_{T,\mu}$ model given in \cite{PCO}. In
section 3, we analyze the effect of chemical potential and
nonconformality on the jet quenching parameter for this model. The
last section is devoted to summary and discussion.

%%%%%%%%%%%%%%%%%%%%%%%%%%%%%%%%%%%%%%%%

\section{Setup}
The SW$_{T,\mu}$ model is defined by the AdS-Reissner Nordstrom
black-hole (AdS-RN) multiplied by a warp factor, given by
\cite{PCO}
\begin{equation}
ds^2=\frac{r^2h(r)}{R^2}(-fdt^2+d\vec{x}^2)+\frac{R^2h(r)}{r^2f}dr^2,\label{metric}
\end{equation}
with
\begin{equation}
f=1-(1+Q^2)(\frac{r_h}{r})^4+Q^2(\frac{r_h}{r})^6,\qquad
h(r)=e^{\frac{c^2R^4}{r^2}},
\end{equation}
where $R$ is the radius of AdS. $Q$ represents the black hole
charge, constrained in $0\leq Q\leq\sqrt{2}$. $r$ denotes the 5th
coordinate with $r=\infty$ the boundary and $r=r_h$ the event
horizon. The $h(r)$ term, characterizing the soft wall model,
distorts the background metric and brings the mass scale $c$ (or
nonconformality), where $c$ is also called the deformation
parameter. Note that here we will not focus on a specific model
with fixed $c$, but rather study the behavior of $\hat{q}$ in a
class of models parametrized by $c$. Because of this, we will make
$c$ dimensionless by normalizing it to fixed $T$ and set $0\leq
c/T\leq2.5$, which is believed to be most relevant for a
comparison with QCD \cite{HLL}.

Moreover, the chemical potential reads
\begin{equation}
\mu=\frac{\sqrt{3}Qr_h}{R^2}.
\end{equation}

The temperature reads
\begin{equation}
T=\frac{r_h}{\pi R^2}(1-\frac{Q^2}{2}).
\end{equation}

\section{jet quenching parameter}
Now we follow the argument in \cite{liu} to investigate the
behavior of the jet quenching parameter for the background metric
(\ref{metric}). In the gravity dual description, $\hat{q}$ can be
computed from light-like adjoint Wilson loops. Specifically, one
considers a null-like rectangular Wilson loop $C$ formed by a
quark-antiquark pair with separation $L$ travelling along
light-cone time duration $L_-$. Under the dipole approximation,
which is valid for small $L$ and $LT<<1$, $\hat{q}$ can be
extracted from the Wilson loop expectation value,
\begin{equation}
<W^A[{\cal C}]> \approx \exp [-\frac{1}{4\sqrt{2}}\hat{q}L_-L^2],
\label{jet}
\end{equation}
where the superscript $A$ represents the adjoint representation.

Using the formulas $<W^A[{\cal C}]>\approx <W^F[{\cal C}]>^2$ and
$<W^F[{\cal C}]>\approx\exp[-S_I]$, one gets
\begin{equation}
\hat{q}=8\sqrt{2}\frac{S_I}{L_-L^2},\label{q}
\end{equation}
with $S_I=S-S_0$, where $S$ is the total energy of the quark
anti-quark pair. $S_0$ denotes the inertial mass of two single
quarks. $S_I$ represents the regulated finite on-shell string
worldsheet action.

To carry on the calculation, one needs to rotate coordinate to
light-cone one, e.g.,
\begin{equation}
dt=\frac{dx^++dx^-}{\sqrt{2}},\qquad
dx_1=\frac{dx^+-dx^-}{\sqrt{2}},
\end{equation}
then metric (\ref{metric}) becomes
\begin{equation}
ds^2=-\frac{r^2h(r)}{R^2}(1+f)dx^+dx^-+\frac{r^2h(r)}{R^2}(dx_2^2+dx_3^2)+\frac{r^2h(r)}{2R^2}(1-f)[(dx^+)^2+(dx^-)^2])+\frac{R^2h(r)}{r^2f}dr^2.
\label{metric1}
\end{equation}

Considering the Wilson loop stretches across e.g., $x_2$ and lies
at $x^+=constant,x_3=constant$, one may choose the following
static gauge
\begin{equation}
x^-=\tau, \qquad x_2=\sigma,
\end{equation}
and assume a profile of $r=r(\sigma)$, then (\ref{metric1})
reduces to
\begin{equation}
ds^2=h(r)[\frac{1}{2}(\frac{r^2}{R^2}-f_1)d\tau^2+(\frac{r^2}{R^2}+\frac{\dot{r}^2}{f_1})d\sigma^2],
\label{metric2}
\end{equation}
with $\dot{r}=\frac{dr}{d\sigma}$, $f_1\equiv\frac{r^2}{R^2}f$.

Given that, the induced metric reads
\begin{equation} g_{00}=\frac{h(r)}{2}(\frac{r^2}{R^2}-f_1), \qquad g_{01}=g_{10}=0, \qquad
g_{11}=h(r)(\frac{r^2}{R^2}+\frac{\dot{r}^2}{f_1}). \label{ind}
\end{equation}

The string is governed by the Nambu-Goto action, given by
\begin{equation}
S=-\frac{1}{2\pi\alpha^\prime}\int d\tau
d\sigma\sqrt{-detg_{\alpha\beta}}, \label{NG}
\end{equation}
with
\begin{equation}
g_{\alpha\beta}=G_{\mu\nu}\frac{\partial
X^\mu}{\partial\sigma^\alpha} \frac{\partial
X^\nu}{\partial\sigma^\beta},
\end{equation}
where $X^\mu$ and $G_{\mu\nu}$ are the target space coordinates
and metric, respectively.

Plugging (\ref{ind}) into (\ref{NG}), one has
\begin{equation}
S=\frac{\sqrt{2}L_-}{2\pi\alpha^\prime}\int_0^{\frac{L}{2}}d\sigma\sqrt{h^2(r)(\frac{r^2}{R^2}-f_1)(\frac{r^2}{R^2}+\frac{\dot{r}^2}{f_1})},\label{NGG}
\end{equation}
where the boundary condition is $r(\pm\frac{L}{2})=\infty$.

As action (\ref{NGG}) does not depend explicitly on $\sigma$, one
gets a conserved quantity
\begin{equation}
\frac{\partial\mathcal L}{\partial\dot{r}}\dot{r}-\mathcal
L=\frac{-h^2(r)(\frac{r^2}{R^2}-f_1)\frac{r^2}{R^2}}{\sqrt{h^2(r)(\frac{r^2}{R^2}-f_1)(\frac{r^2}{R^2}+\frac{\dot{r}^2}{f_1})}}=C,
\end{equation}
results in
\begin{equation}
\dot{r}^2=\frac{f_1r^2}{R^2C^2}[\frac{h^2(r)
r^2(\frac{r^2}{R^2}-f_1)}{R^2}-C^2].\label{rr}
\end{equation}

The above equation involves determining the zeros. Also, the
turning point occurs at $f_1=0$, indicating $\dot{r}=0$ at $r=r_h$
\cite{liu}.

For convenience, we write $B\equiv1/C^2$. For $C\rightarrow0$ (the
low energy limit), one can integrate (\ref{rr}) to leading order
in $1/B$, yielding
\begin{equation}
L=2R^2\int_{r_t}^\infty dr
\sqrt{\frac{1}{(\frac{r^2}{R^2}-f_1)Bf_1r^4h^2(r)}}.\label{LL}
\end{equation}

Putting (\ref{rr}) into (\ref{NGG}), one gets
\begin{eqnarray}
S&=&\frac{\sqrt{2}L_-}{2\pi\alpha^\prime}\int_{r_h}^\infty dr
\sqrt{\frac{h^4(r)(\frac{r^2}{R^2}-f_1)^2r^2}{f_1[h^2(r)
r^2(\frac{r^2}{R^2}-f_1)-R^2C^2]}}\nonumber\\
&=&\frac{\sqrt{2}L_-\sqrt{B}}{2\pi\alpha^\prime}\int_{r_h}^\infty
dr\frac{h^2(r)(\frac{r^2}{R^2}-f_1)r}{\sqrt{h^2(r)(\frac{r^2}{R^2}-f_1)Bf_1r^2-f_1R^2}}.\label{NGG1}
\end{eqnarray}

Similarly, one expands (\ref{NGG1}) to leading order in $1/B$ as,
\begin{equation}
S=\frac{\sqrt{2}L_-}{2\pi\alpha^\prime}\int_{r_h}^\infty
dr[1+\frac{R^2}{2h^2(r)(\frac{r^2}{R^2}-f_1)Br^2}]
\sqrt{\frac{1}{f_1}h^2(r)(\frac{r^2}{R^2}-f_1)}. \label{SS}
\end{equation}

However, action (\ref{SS}) is divergent. To eliminate the
divergence it should be subtracted by the inertial mass of two
single quarks, given by
\begin{eqnarray}
S_0&=&\frac{2L_-}{2\pi\alpha^\prime}\int_{r_h}^\infty
dr\sqrt{g_{--}g_{rr}}\nonumber\\
&=&\frac{\sqrt{2}L_-}{2\pi\alpha^\prime}\int_{r_h}^\infty dr
\sqrt{\frac{1}{f_1}h^2(r)(\frac{r^2}{R^2}-f_1)}.
\end{eqnarray}

Then the regulated finite on-shell action is given by
\begin{equation}
S_I=S-S_0=\frac{\sqrt{2}L_-R^2}{4\pi\alpha^\prime
B}\int_{r_h}^\infty
dr\sqrt{\frac{1}{(\frac{r^2}{R^2}-f_1)f_1r^4h^2(r)}}.\label{SI}
\end{equation}

Substituting (\ref{LL}) and (\ref{SI}) into (\ref{q}), one ends up
with the jet quenching parameter in the SW$_{T,\mu}$ model
\begin{equation}
\hat{q}=\frac{I(q)^{-1}}{\pi\alpha^\prime},\label{q1}
\end{equation}
with
\begin{equation}
I(q)=R^2\int_{r_h}^\infty
dr\sqrt{\frac{1}{(\frac{r^2}{R^2}-f_1)f_1r^4h^2(r)}}.
\end{equation}

Note that by setting $c=\mu=0$ in (\ref{q1}), the jet quenching
parameter of SYM \cite{liu} is reproduced, that is
\begin{equation}
\hat{q}_{SYM}=\frac{\pi^{3/2}\Gamma(\frac{3}{4})}{\Gamma(\frac{5}{4})}\sqrt{\lambda}T^3,
\end{equation}
where one has used the relations $r_h=\pi R^2T$ and
$\frac{R^2}{\alpha^\prime}=\sqrt{\lambda}$.

\begin{figure}
\centering
\includegraphics[width=8cm]{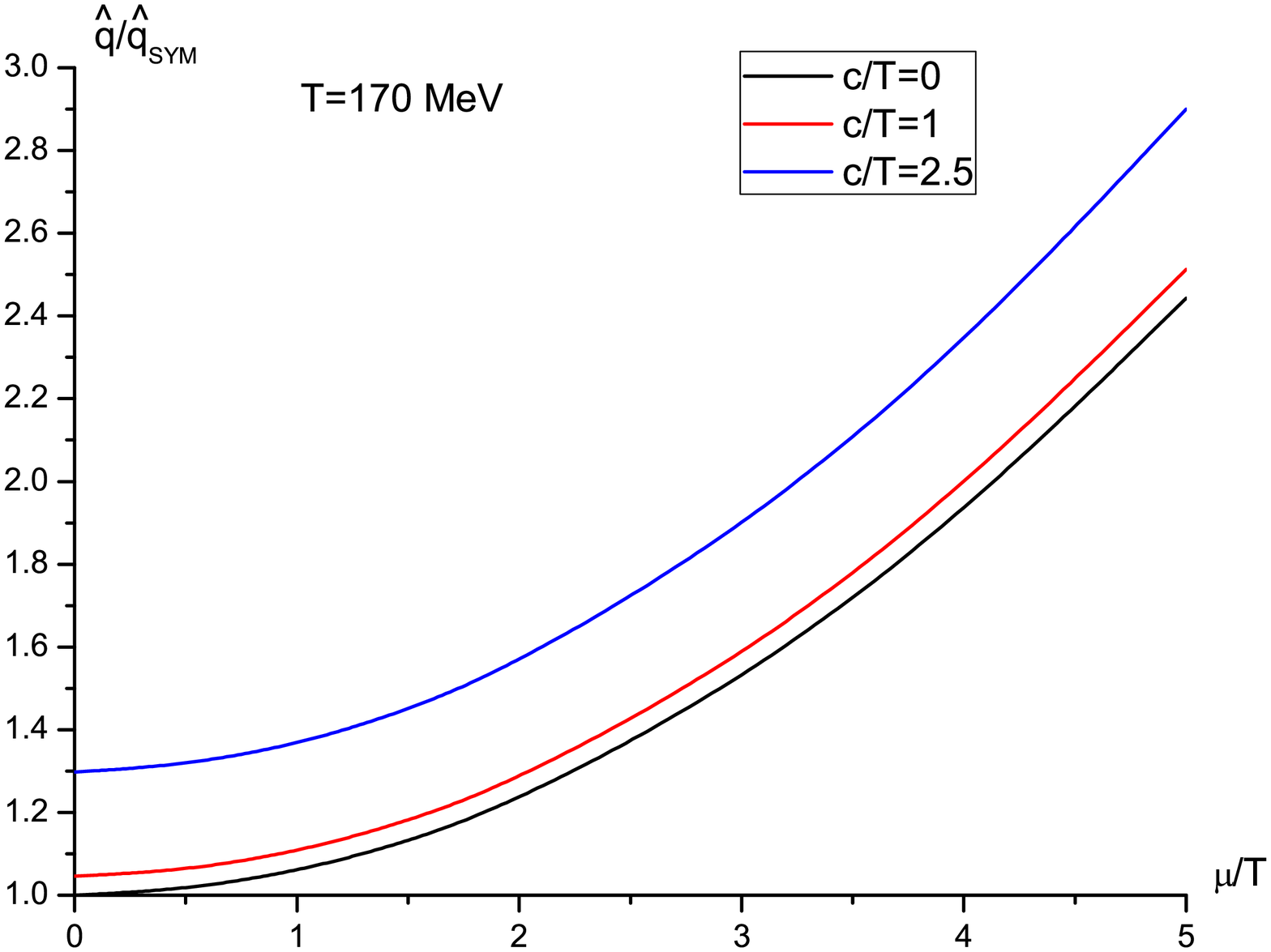}
\includegraphics[width=8cm]{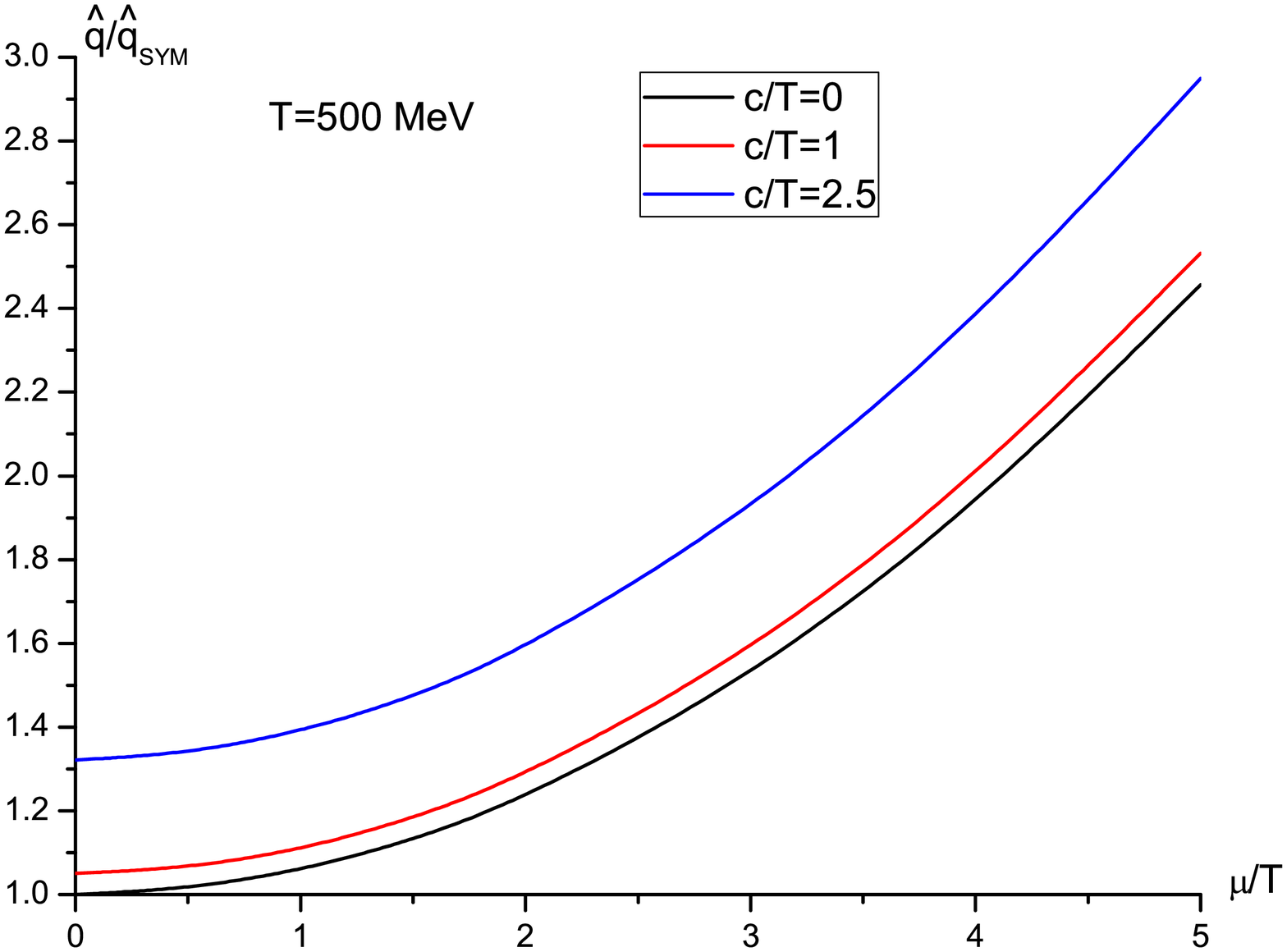}
\caption{$\hat{q}/\hat{q}_{SYM}$ versus $\mu/T$ with fixed $c/T$.
Left: $T=170 MeV$. Right: $T=500 MeV$. In both panels from top to
bottom, $c/T=2.5, 1, 0$, respectively.}
\end{figure}

\begin{figure}
\centering
\includegraphics[width=8cm]{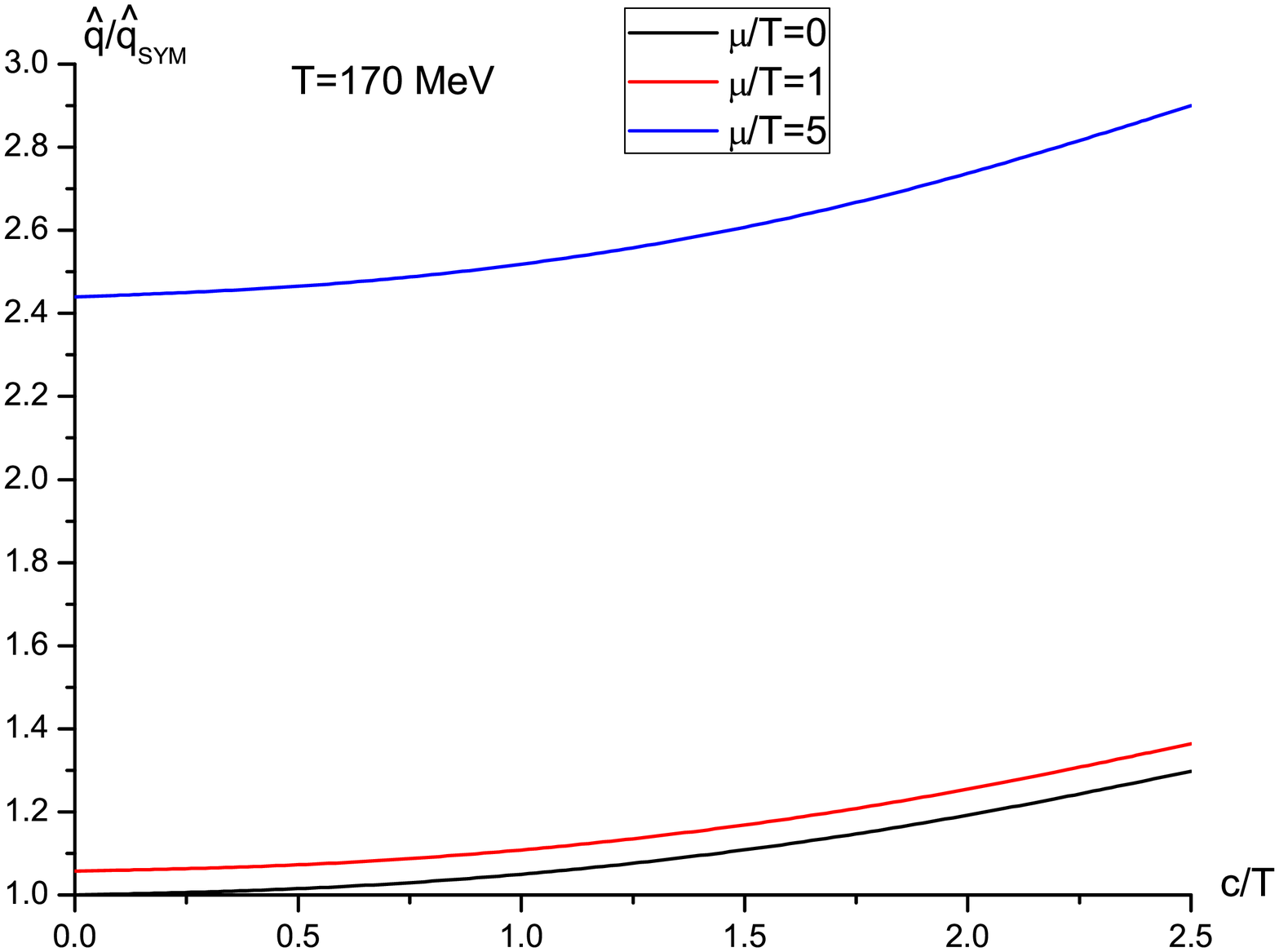}
\includegraphics[width=8cm]{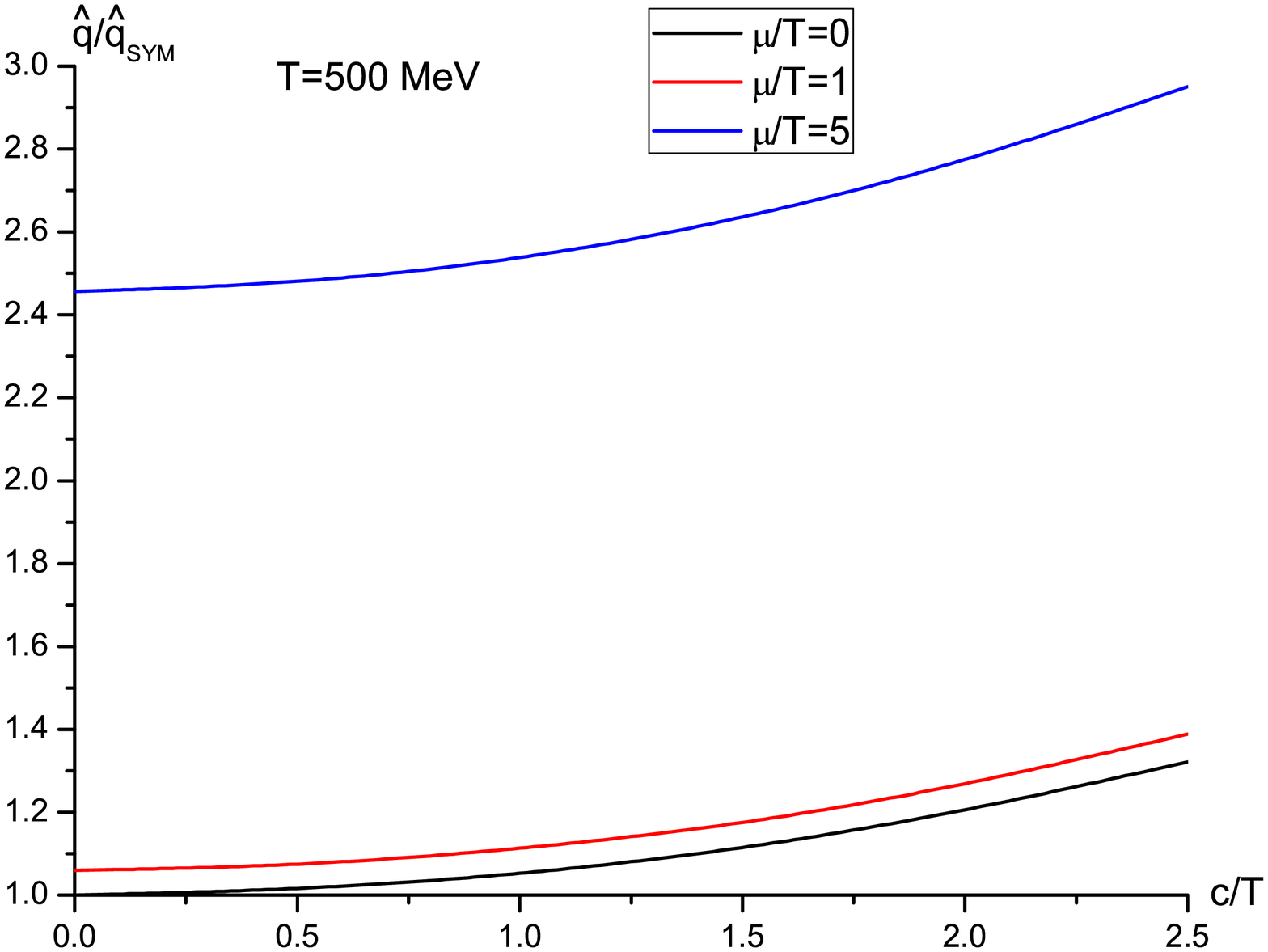}
\caption{$\hat{q}/\hat{q}_{SYM}$ versus $c/T$ with fixed $\mu/T$.
Left: $T=170 MeV$. Right: $T=500 MeV$. In both panels from top to
bottom, $\mu/T=5, 1, 0$, respectively.}
\end{figure}

\begin{figure}
\centering
\includegraphics[width=8cm]{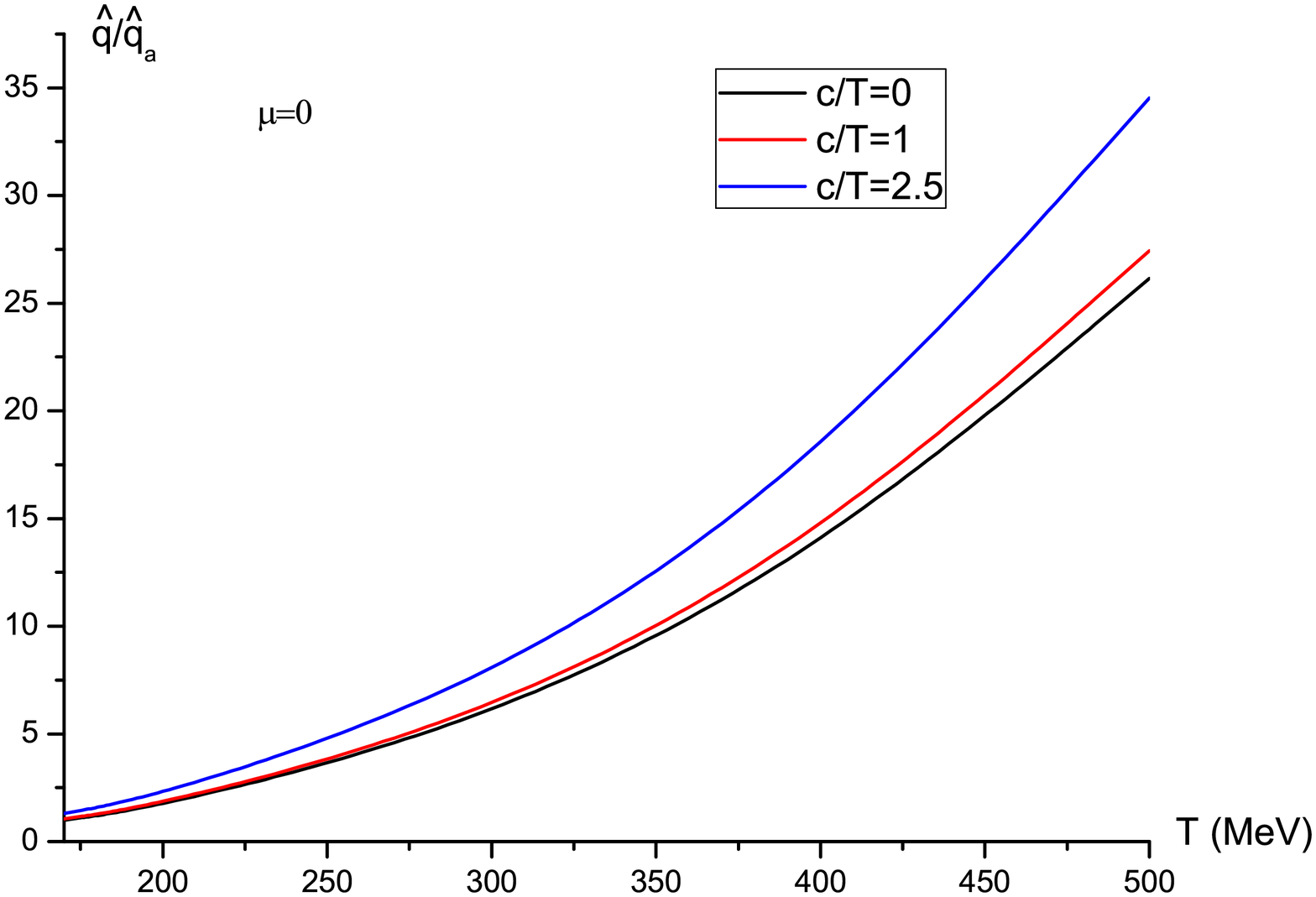}
\includegraphics[width=8cm]{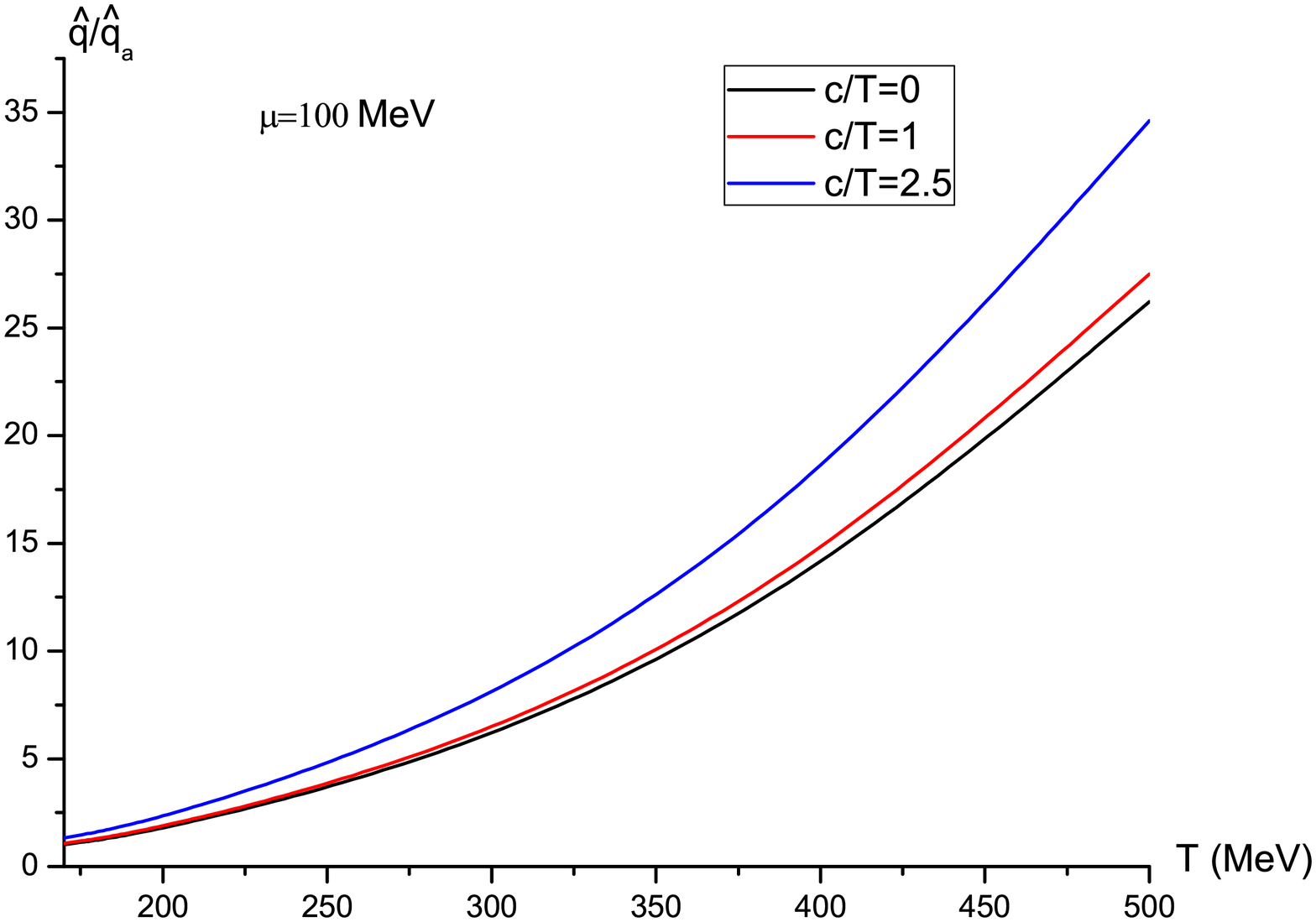}
\caption{$\hat{q}/\hat{q}_a$ versus $T$ with fixed $c/T$. Left:
$\mu=0$. Right: $\mu=100 MeV$. In both panels from top to bottom,
$c/T=2.5, 1, 0$, respectively.}
\end{figure}

Let's discuss results. First, we analyze how $\mu$ and $c$ modify
$\hat{q}$. For this purpose, we plot $\hat{q}/\hat{q}_{SYM}$ as a
function of $\mu/T$ with fixed $c/T$ for two different
temperatures in Fig.1, where the left panel is for $T=170 MeV$
while the right $T=500 MeV$. From both panels, one sees at fixed
$c/T$, increasing $\mu/T$ leads to increasing
$\hat{q}/\hat{q}_{SYM}$, indicating the inclusion of chemical
potential increases the jet quenching parameter, in accord with
that found in \cite{FL,SD}. Likewise, one can see from Fig.2 that
at fixed $\mu/T$, $\hat{q}/\hat{q}_{SYM}$ increases as $c/T$
increases, implying the inclusion of nonconformality increases the
jet quenching parameter, similar to \cite{AS}. Thus, one concludes
that the inclusion of chemical potential and nonconformality both
increase the jet quenching parameter thus enhancing the energy
loss, consistently with the findings of the drag force \cite{YH}.

Also, we want to understand the $T$ dependence of $\hat{q}$ for
this model. To this end, we plot $\hat{q}/\hat{q}_a$, with
$\hat{q}_a|_{c=\mu=0, T=170 MeV}$, versus $T$ in Fig.3, where the
left panel is for $\mu=0$ while the right $\mu=100 MeV$. From
these figures, one finds with fixed $c/T$, $\hat{q}/\hat{q}_a$
increases as $T$ increases, as expected.

Finally, we would like to make a comparison to implications of
experiment data. In Tab. 1, we present some typical values of
$\hat{q}$, where we have taken $N_c=3$ and $\alpha_{SYM}=0.5$
(reasonable for temperatures not far above QCD phase transition),
and $\lambda=6\pi$ \cite{liu}. One finds that most of the values
are consistent with the extracted values from RHIC data
($5\sim25GeV ^2/fm$) \cite{KFE,JD}. On the other hand, since the
presence of $\mu$ and $c$ both enhance the jet quenching
parameter, one may infer that increase $\mu$ and $c$ may lower the
possible allowed domain of $T$ for the computed $\hat{q}$ to agree
with the experiment data.

\begin{table}[hbtp]
\centering
\begin{tabular}{|c|c|c|c|c|c|c|c|c|c|}\hline
$T\setminus(\mu, c)$&(0, 0)&(0, 0.3)&(0, 0.7)&(0.1, 0)&(0.1,
0.3)&(0.1, 0.7)&(0.3, 0)&(0.3, 0.3)&(0.3, 0.7)
\\ \hline
$0.3$&4.50&4.71&5.70&4.53&4.74&5.73&4.76&4.98&6.0
\\ \hline
$0.4$&10.61&10.89&12.19&10.64&10.93&12.23&10.94&11.23&12.56
\\ \hline
$0.5$&20.69&21.02&22.65&20.70&21.06&22.70&21.08&21.45&23.10
\\ \hline
\multicolumn{3}{c}{}
\end{tabular}
\label{exampletable} \caption{Typical values of $\hat{q}$ in $GeV
^2/fm$, where the first line denotes $(\mu, c)$ and the first
column indicates $T$. Here $T,\mu,c$ are all expressed in units of
$GeV$.}
\end{table}

\section{Conclusion}
In this paper, we studied the jet quenching parameter in a soft
wall model with finite temperature and chemical potential. The
dual space geometry is AdS-RN black hole (describe finite
temperature and density in the boundary theory) multiplied by a
background warp factor (generate confinement). Our motivation
rests on the earlier studies of the free energy \cite{PCO},
imaginary potential \cite{zq} and drag force \cite{YH} in such a
model. It turns out that the inclusion of chemical potential and
nonconformality both increase the jet quenching parameter thus
enhancing the energy loss, in agreement with the findings of the
drag force \cite{YH}. Also, we attempted to make a comparison to
implications of experiment data and found the theoretical
estimates agree well with experiment results. Finally, our results
suggested that increase $\mu$ and $c$ may lower the possible
allowed domain of $T$ for the computed $\hat{q}$ to agree with the
experiment data.

Admittedly, the SW$_{T,\mu}$ model has some drawbacks. The primary
disadvantage is that it is not a consistent model since it doesn't
solve the Einstein equations. Studying the jet quenching parameter
in some consistent models, e.g. \cite{SH,SH1,DL0,RCR} would be
informative (but note that the metrics of those models are only
known numerically, so the calculations are very complex).
Moreover, the SW$_{T,\mu}$ model may miss a part about the phase
transition \cite{DL1,RG2,FZ} and the effect of non-trivial dilaton
field \cite{UG1,JNO,MM}). Considering these effects would also be
instructive. These will be left for further studies.

\section{Acknowledgments}
This work is supported by the NSFC under  Grants Nos. 11705166,
11947410, and Zhejiang Provincial Natural Science Foundation of
China No. LY19A050001 and No. LY18A050002.

%%%%%%%%%%%%%%%%%%%%%%%%%%%%%%%%%%%%%%%%

\end{document}